\documentstyle[12pt]{article}
\normalbaselines
\begin{document}

\begin{center}
{\Large\bf Probability Distributions and Hilbert Spaces: Quantum
and Classical Systems}\\[7mm]
{V. I. Man'ko\footnote{on leave from the P. N. Lebedev
Physical Institute, Moscow, Russia} and G. Marmo}\\
{\it Dipartimento di Scienze Fisiche, Universit\`a di Napoli
``Federico II'' and\\
Istituto Nazionale di Fisica Nucleare, Sezione di
Napoli, Mostra d'Oltremare, Pad. 20, 80125 Napoli, Italia}\\[5mm]
\end{center}

\vspace{2mm}

\begin{abstract}
We use the fact that some linear Hamiltonian systems can be considered
as ``finite level'' quantum systems,
and the description of quantum mechanics in terms of probabilities, to
associate probability distributions with this particular class of linear 
Hamiltonian systems. 
\end{abstract}

\section{Introduction}

\noindent

Some general structural features of dynamical systems can be exhibited in 
a study of the relation between classical and quantum mechanics. The most 
widely accepted formal relationship between quantum mechanics and classical 
mechanics is expressed in the analogy between commutator brackets of operators 
on some Hilbert space and  Poisson brackets among functions on some
symplectic phase space, along with the analogy between Heisenberg equations
of motion and Hamilton equations of motion. This analogy can be made more
suggestive if operators are replaced by expectation values (quadratic 
functions on the Hilbert space) and the commutator is replaced by Poisson
brackets among these quadratic functions~\cite{Bregenz97}. 
It is also possible to
show that Schr\"odinger equations define  Hamiltonian systems on the 
Hilbert space, where the imaginary part of the scalar product defines a 
symplectic structure. These various aspects have been considered
in the past~\cite{Strocchi} and more recently in connection with Wigner's
problem~\cite{Wignerpaper,Ramon}. They can be summarized in the following way.

On a symplectic vector space, carrier space of a Hamiltonian linear dynamical
system, one introduces a complex structure compatible with the symplectic
structure, i.e., their composition, as maps, defines a positive definite 
inner product on the vector space. Finite-level quantum systems are shown to
be linear Hamiltonian systems which, in addition, leave invariant a compatible
complex structure. Therefore, on a finite-dimensional Hilbert space quantum
systems are a subclass of linear Hamiltonian systems. This characterization
carries over to infinite dimensions where the ``classical system'' has an
infinite numbers of degrees of freedom. By no means, the linear system 
preserving two compatible structures (quantum system) can
be considered as a quantization of the associated Hamiltonian one (classical 
system).

Continuing in the spirit of the approach by Moyal~\cite{Moyal49}, 
Wigner~\cite{Wigner32}, Husimi~\cite{Husimi40}, et al., 
according 
to which quantum mechanics can be formulated in a natural manner in terms of
functions on the classical phase space, recently~\cite{Tombesi2} a different 
approach 
has been proposed dealing with probability distributions rather than 
quasi-probabilities of previous approaches. It is to be remarked that this
approach differs from the one initiated by Koopman~\cite{Koopman}, who had shown how the
dynamical transformations of classical mechanics considered as measure
preserving transformations of the phase space, induce unitary transformations 
on the Hilbert space of functions which are square integrable with respect to 
a Liouville measure over the phase space. It is to be stressed that this Hilbert
space corresponds not to the space of state vectors in quantum mechanics but 
to the Hilbert space of operators on state vectors (with the trace of 
the product of two operators being chosen as the scalar product).

In this paper, we would like to consider classical canonical linear systems
identified with the ``Hamiltonian part'' of quantum systems with a finite 
number of states. We use previous correspondence with probabilities, to
associate probabilities also with classical linear systems. Then we move to 
infinite dimensions and show how things appear there. This study shows the
main difference between classical systems and quantum systems at the level 
of associated probabilities.

In a future paper, we shall consider how the dynamical evolution and 
dynamical symmetry groups are represented on these probability 
distributions.

The state of a classical particle (one degree of freedom) is described 
by two $c$-number valued functions, position $q$ and momentum 
$p$~\cite{Landau}. In
classical statistical mechanics, the state of a particle is described 
by the probability distribution on the particle's phase space~\cite{Landau}
$f(q,p)\geq 0$ such that
\begin{equation}\label{H1}
\int f(q,p)\,dq\,dp=1\,.
\end{equation}
In quantum mechanics, pure states of a particle are described by a
complex wave function $\Psi (q)$ and mixed normalized states are described
by density matrices, $\rho (q,q')$ is the matrix element 
$\langle q\mid \hat \rho \mid q'\rangle $
of an Hermitian nonnegative density operator $\hat \rho $ with 
$\mbox {Tr}\,\hat \rho=1.$ The diagonal elements of the density matrix 
$\rho (q,q)$ determine the probability distribution in position which
for pure state $\hat \rho _\Psi =\mid \Psi \rangle \langle \Psi \mid $ 
is reduced to $\rho _\Psi (q,q)=|\Psi (q)|^2.$ The wave function $\Psi (q)$
may be considered as a component $\Psi _q\equiv \Psi (q)$ of the state vector
in Hilbert space. The Hilbert space structure is usually considered as 
appropriate for description of the quantum motion. 

On the other hand,
Koopman~\cite{Koopman} has demonstrated that a Hilbert space structure
can be associated to a classical system  as well. This approach
has been developed in~\cite{Marmothesis}. Also Strocchi~\cite{Strocchi}
has shown that  quantum motion (Schr\"odinger 
equation~\cite{Schr26} for the wave function) may be reformulated as
formally classical motion of coupled harmonic oscillators.
The linear dynamics of such oscillators and its possible deformations
were discussed in~\cite{Vietri,Wignerpaper,Bregenz97}. 

The relation of quantum dynamics to a probability distribution evolution
(probability representation of quantum mechanics) was found 
in~\cite{Tombesi1} in connection with tomography 
transformation~\cite{WVogel,Tombesi3} (Radon transform) of the density 
matrix used for measuring quantum states~\cite{Raymer,Wunsche}.
In probability representation, a quantum state is described by a family
of positive probability distributions of position instead of wave function
or density matrix. Due to this the probability representation can be 
considered as classical-like formulation of quantum mechanics. The connection
of irreducible representations of Heisenberg--Weyl group with the tomography
transformation was discussed in~\cite{MankoMarmo}.

The goal of our paper is to clarify some mutual relations between these 
different aspects of  classical and quantum dynamics. We demonstrate
that the classical motion of a system, both with finite and infinite numbers
of degrees of freedom, can be associated with the probability measure
(a family of marginal distributions) characteristic for quantum mechanics.

The paper is organized as follows. In Section~2, we review description
of quantum dynamics in terms of classical formalism. In Section~3, we 
consider two-level system in detail. Section~4 is devoted to treating
two-level system as a spin-1/2-particle system. In Section~5, the
probability representation (tomographic map) of quantum mechanics of
spin-1/2 particle is reviewed. In Section~6, the map of states of two 
and $N$ classical oscillators onto probability distributions is considered.
In Section~7, we discuss classical states of infinite number of oscillators
in terms of the marginal distributions and properties of the probability
distributions under action of canonical transform of the oscillators' 
phase space.

\section{Classical Formalism for Quantum Dynamics}

\noindent

The Schr\"odinger evolution equation $(\hbar =1)$ for the wave function 
of a one-dimensional particle
\begin{equation}\label{H2}
i\dot \Psi(q,t)=H\Psi (q,t)\,,
\end{equation}
where the Hamiltonian operator is
\begin{equation}\label{H3}
H=\frac {\hat p^2}{2}+V(\hat q)\,,\qquad 
\hat p=-i\frac {\partial}{\partial q}\,,\qquad \hat q=q\,,
\end{equation}
in the coordinate representation for real and imaginary parts of 
the wave function 
\begin{equation}\label{H4}
\Psi(q,t)\equiv \Psi _q=\frac {1}{\sqrt 2}\left(x_q+ip_q\right);
\qquad 
\Psi ^*(q,t)\equiv \Psi _q^*=\frac {1}{\sqrt 2}\left(x_q-ip_q\right)
\end{equation}
can be rewritten in the form~\cite{Strocchi}
\begin{equation}\label{H5}
i\frac {\partial}{\partial t}\left(\begin{array}{c}
\Psi _q\\
\Psi _q^*\end{array}\right)=\left(\begin{array}{clcr}
H_{qq'}&0\\
0&-H_{qq'}^*\end{array}\right)\left(\begin{array}{c}
\Psi _{q'}\\
\Psi _{q'}^*\end{array}\right).
\end{equation}
We treat $H_{qq'}$ as matrix elements of the Hamiltonian in the coordinate
representation
\begin{equation}\label{H6}
\langle q\mid H\mid q'\rangle =H_{qq'}
\end{equation}
and perform summation (integration) over repeated indices $q'.$ By introducing 
notation
\begin{equation}\label{H7}
\xi _q=\left(\begin{array}{c}
\Psi _q\\
\Psi _q^*\end{array}\right),
\end{equation}
we rewrite the Schr\"odinger equation in the form
\begin{equation}\label{H8}
i\dot \xi=h_{qq'}\xi _{q'}\,, \qquad
h_{qq'}=\left(\begin{array}{clcr}
H_{qq'}&0\\
0&-H_{qq'}^*\end{array}\right).
\end{equation}
The evolution equation can be obtained as a classical equation using the
Hamiltonian
\begin{equation}\label{H9}
{\cal H}=\Psi_q^*
H_{qq'}\Psi_{q'}
\end{equation}
with summation (integration) over repeated indices and with Poisson brackets
\begin{equation}\label{H10}
\left \{\Psi _q, \Psi _{q'}\right \}=\left \{\Psi _q^*, 
\Psi _{q'}^*\right \}=0\,; \qquad
\left \{\Psi _q, \Psi _{q'}^*\right \}=i\delta _{qq'}\,.
\end{equation}
The symbol $\delta _{qq'}$ is either Kronecker symbol (if $q$ and $q'$ are
discrete indices) or Dirac delta-function (for continuous indices $q$ and 
$q'$). Thus, Schr\"odinger equation is equivalent to the Hamiltonian 
``classical equation''
\begin{equation}\label{H11}
\dot \Psi_q=\left \{{\cal H},\Psi _q\right \}.
\end{equation}
One can see that Schr\"odinger equations are equivalent to classical equations
for a system of coupled oscillators, with coupling of the oscillators being
described by matrix elements $H_{qq'}.$ To show this more explicitly, we 
rewrite the equations for real variables $x_q, p_q,$ for which Poisson brackets
are real ones
\begin{equation}\label{H12}
\left \{x_q, x_{q'}\right \}=\left \{p_q, 
p_{q'}\right \}=0\,; \qquad
\left \{p_q, x_{q'}\right \}=i\delta _{qq'}\,.
\end{equation}
Then the Hamiltonian takes the form
\begin{equation}\label{H13}
{\cal H}=\frac {1}{2}Q_qB_{qq'}Q_{q'}\,,
\end{equation}
where
\begin{equation}\label{H14}
Q_q=\left(\begin{array}{c}
p_q\\
x_q\end{array}\right) \qquad \mbox {and} \qquad
B_{qq'}=\frac {1}{2}\left(\begin{array}{clcr}
H_{qq'}+H_{qq'}^*&i\left(H_{qq'}^*-H_{qq'}\right)\\
i\left(H_{qq'}-H_{qq'}^*\right)&H_{qq'}+H_{qq'}^*
\end{array}\right).
\end{equation}
Then the evolution equation for the real vector $Q_q$ reads
\begin{equation}\label{H15}
\dot Q_q=A_{qq'}Q_{q'}\,,
\end{equation}
with summation (integration) over $q'$ and the matrix $A_{qq'}$
\begin{equation}\label{H16}
A_{qq'}=-\Sigma B_{qq'}\,;\qquad \Sigma =\left(\begin{array}{clcr}
0&1\\
-1&0\end{array}\right).
\end{equation}
We see that the search for the energy levels of the quantum system is 
equivalent to the search for normal modes and their frequencies for a
corresponding classical system of coupled oscillators.

\section{Two-Level System}

\noindent

Till now we used infinite system of coupled oscillators and our manipulations
were done at a formal level. To clarify the
approach for constructing probability distribution associated to a classical
state, let us consider a two-level system. In this case, 
following~\cite{Wignerpaper} we have
\begin{equation}\label{H17}
\Psi=\left (\begin{array}{c}\Psi_1 \\ \Psi_2\end{array}\right )
\qquad \mbox {and}
\qquad \xi=\left (\begin{array}{c}\Psi \\ \Psi^*\end{array}\right ).
\end{equation}
The Hermitian matrix $H$ such that 
\begin{equation}\label{H18}
H_{ik}=H_{ki}^*=\left(\begin{array}{clcr}
a&b_1+ib_2\\
b_1-ib_2&c\end{array}\right);\qquad i,k=1,2
\end{equation}
is the Hamiltonian and Schr\"odinger equation looks as
\begin{equation}\label{H19}
i\dot \xi =h\xi;\qquad h==\left(\begin{array}{clcr}
H&0\\
0&-H^*\end{array}\right).
\end{equation}
Energy levels of a two-level system are determined by matrix elements
$a,b=b_1+ib_2,c$ of matrix~(\ref{H18}) and we arrive at 
\begin{eqnarray}
E_1&=&\frac {a+c}{2}+\frac {1}{2}\sqrt {(a+c)^2+4(bb^*-ac)}\,;\nonumber\\
\label{w1}
E_2&=&\frac {a+c}{2}-\frac {1}{2}\sqrt {(a+c)^2+4(bb^*-ac)}\,.\nonumber
\end{eqnarray}
By introducing the real components of the wave function
\begin{equation}\label{H20}
\Psi _k=\frac {1}{\sqrt 2}\left(x_k+ip_k\right),
\end{equation}
i.e.,
\begin{equation}\label{H21}
\mbox {\bf x}=\left (\begin{array}{c}x_1 \\ x_2\end{array}\right );\qquad 
\mbox {\bf p}=\left (\begin{array}{c}p_1 \\ p_2\end{array}\right ),
\end{equation}
and
\begin{equation}\label{H22}
\mbox {\bf Q}=\left (\begin{array}{c}\mbox {\bf p}\\ 
\mbox {\bf x}\end{array}\right ),
\end{equation}
we arrive at the Hamiltonian description of two formally classical 
coupled oscillators in the form
\begin{equation}\label{H23}
{\cal H}=\frac {1}{2}\sum _{\alpha,\beta=1}^4Q_\alpha B_{\alpha \beta}
Q_\beta\,,
\end{equation}
where 4$\times $4-matrix $B$ reads
\begin{equation}\label{H24}
B=\frac {1}{2}\left(\begin{array}{clcr}
H+H^*&i\left(H^*-H\right)\\
-i\left(H-H^*\right)&H+H^*
\end{array}\right).
\end{equation}
The Schr\"odinger evolution equation for the two-level system reads
\begin{equation}\label{H25}
\dot {\mbox {\bf Q}}=A\mbox {\bf Q}\,,
\end{equation}
where
$$A=-\Sigma B\,.$$
Thus, the quantum linear dynamics of two-level system is equivalent to
the classical dynamics of two coupled classical oscillators. This means
that one can associate to two coupled classical oscillators the 
two-dimensional Hilbert space of states which formally is equivalent to 
the artificial spin-1/2 states. 

The classical states of two oscillators 
is described by four real numbers $q_1,p_1,q_2,p_2.$
The Hilbert structure is introduced by relating these four numbers
to ``spinors'' 
\begin{equation}\label{H26}
\Psi =\frac {1}{\sqrt 2}\left(\begin{array}{c}
q_1+ip_1\\
q_2+ip_2\end{array}\right).
\end{equation}
On the other hand, the distribution functions which can determine completely 
the spinors were considered in~\cite{OlgaJETF}.
Thus, we can associate with two linear classical coupled oscillators a
positive probability distributions (see below).

\section{Two-Level System as Spin-1/2 System}

\noindent

Let us describe now a two-level quantum system and its states in terms of
spinorial states. To proceed in this direction, we construct the spinor $\Psi $
as a column consisting of two-state vectors (of the two-level system) $\Psi_1$
and $\Psi _2$ as given by formula~(\ref{H17}).
Then the density matrix of the two-level system can be constructed for 
the pure states as
\begin{equation}\label{H28}
\rho _{ik}^{(\Psi)}=\Psi _i\Psi _k^*;\qquad i,k=1,2.
\end{equation}
For mixed state, the density operator has the form
\begin{equation}\label{H29}
\hat \rho =w_1\mid \Psi _1\rangle\langle \Psi_1\mid
+w_2\mid \Psi _2\rangle\langle \Psi_2\mid\,,
\end{equation}
where $w_1$ and $w_2$ are positive probabilities and $w_1+w_2=1.$
The generic Hermitian density matrix of the mixed state of the spin-1/2 
\begin{equation}\label{H30}
\rho _{ik}=\left(\begin{array}{clcr}
\rho _{11}&\rho _{12}\\
\rho _{12}^*&\rho _{22}
\end{array}\right)
\end{equation}
has the property $\rho _{11}+\rho _{22}=1.$

For pure state, we have
\begin{equation}\label{H31}
\rho _{11}^{(\Psi)}=|\Psi _1|^2;\qquad \rho _{22}^{(\Psi)}=|\Psi _2|^2;
\qquad \rho _{12}^{(\Psi)}=\Psi _1\Psi _2^.
\end{equation}
Thus, we can associate to classical two-mode oscillator and to its states
(with positions $q_1,q_2$ and momenta $p_1,p_2$) the density matrix of
spin-1/2 particle.

\section{Tomographic map for spinors}

\noindent

Tomography of spin states was considered in~\cite{OlgaJETF,DodMan}. The
marginal distribution which determines the spin state can be constructed
using the relation
\begin{equation}\label{H32}
\rho _{mm'}(\varphi, \theta ,\psi)=({\cal D}\rho {\cal D}^\dagger)_{mm'}\,.
\end{equation}
Here $\rho$ is a density matrix in the initial reference frame and 
$\rho _{mm'}(\varphi, \theta)$ is the density matrix in the rotated 
reference frame with Euler angles $(\varphi, \theta ,\psi).$
The matrix ${\cal D}$ is the matrix of the spinor irreducible representation
of the group $SU(2)$ and its matrix elements depend on the Euler
angles $(\varphi, \theta ,\psi).$ In view of the structure of the formula
for rotated density matrix, the diagonal elements of the density matrix 
depend on two angles $(\varphi, \theta )$ only determining the direction 
of the quantization axis. 
The one-dimensional stability group of $\rho$ in $SU(2)$ determines the
quotient $S^2=SU(2)/U(1).$
The spin projection on this axis has the 
probability distribution function
\begin{equation}\label{H33}
w\left(m,\varphi,\theta\right)=\rho _{mm}(\varphi, \theta ).
\end{equation}
The function $w\left(m,\varphi,\theta\right)$ is positive 
and normalized
$$
\sum_{m=-1/2}^{1/2}w\left(m,\varphi,\theta\right)=1\,.
$$
Since the probability distribution function depends on two extra parameters
$\left(\varphi,\theta\right)$, it is possible to express both diagonal and
nondiagonal elements of the initial density matrix which determines the
spin quantum state in terms of the distribution function~\cite{OlgaJETF}\,:
\begin{eqnarray}\label{H34}
&\,&(-1)^{m_2'}\sum_{j_3=0}^1\,\sum_{m_3=-j_3}^{j_3}\,(2j_3+1)^2\nonumber\\
&\times&\sum_{m_1
=-1/2}^{1/2} \int (-1)^{m_1} w\left(m_1,\varphi,\theta\right)\, 
D_{0 m_3}^{(j_3)}
(\varphi,\theta,\psi)\nonumber\\
&\times& \pmatrix{j&j&j_3\cr m_1&-m_1&0}\pmatrix{j&j&j_3\cr m_1'&-m_2'&m_3}
\frac{d\omega}{8\pi^2}=\rho_{m_1'm_2'}^{(j)},  \end{eqnarray}
where
\begin{equation}\label{Zheq.12}
 \int d\omega =\int _0^{2\pi}d\varphi \int _0^{\pi} \sin \,\theta \,d\theta
\int _0^{2\pi}d\psi \,.
\end{equation}
In formula~(\ref{H34}), Wigner $3j$-symbols and matrix elements of the 
irreducible representation of the $SU(2)$ group ${\cal D}_{0m}^{(j_3)}$ are
adopted from~\cite{Landau}. Thus, the states of spin-1/2 particle are mapped
onto set of marginal probability distributios 
$w\left(m,\varphi,\theta\right).$

\section{Map of Classical States of Two-Mode Harmonic Oscillators onto
Probability Distribution Set}

\noindent

In the previous sections, we showed that the states of a classical two-mode 
oscillator labelled by four real $c$-number functions $q_1,q_2,p_1,p_2$ 
are mapped onto state vectors of two-dimensional Hilbert space of spinors
$$q_1,q_2,p_1,p_2\longleftrightarrow \Psi\,.$$
Spinors, which differ by a common phase factor, i.e., $\Psi$ and 
$e^{i\varphi}\Psi,$ determine the same quantum state. Thus, we arrive at
description of pure quantum state of spin-1/2 in terms of nonnegative
Hermitian 2$\times$2 density matrix $\rho $ with trace equal to unity
and condition
$$\mbox {Tr}\,\rho ^2=1\,.$$
This means that states of classical two-mode oscillators are mapped onto
a set of density matrices of the pure spin states,
$$q_1,q_2,p_1,p_2\rightarrow \rho\,.$$
In Section~5, we have shown that the density matrix of spin state determines
the probability distribution function
$$\rho\rightarrow w\left(m,\varphi,\theta\right)$$
of the spin projection $m$ onto the axis with direction given by angles
$\varphi$ and $\theta.$ Also, given the distribution function 
$w\left(m,\varphi,\theta\right)$ one determines the density matrix
$\rho $ due to formula~(\ref{H34})\,:
$$w\left(m,\varphi,\theta\right)\rightarrow \rho\,.$$
Thus, we have proved that the classical state of a two-mode oscillator, 
labelled
by four real numbers $q_1,q_2,p_1,p_2$ can be associated with probability
distribution,
$$q_1,q_2,p_1,p_2\rightarrow w\left(m,\varphi,\theta\right).$$
For the spin $j,$ the construction is similar and formulae for spin-1/2
case are modified as follows
\begin{equation}\label{H35}
\Psi ^{(j)}=\left(\begin{array}{c}
\Psi_1\\
\vdots\\
\Psi_{2j+1}\end{array}\right).
\end{equation}
The density matrix of pure state is given by 
\begin{equation}\label{H36}
\rho _{ik}^{(j)}=\Psi _i\Psi _k^*,\qquad i,k=1,2,\ldots,2j+1\,.
\end{equation}
The density operator of a mixed state has the form
\begin{equation}\label{H37}
\hat \rho ^{(j)}=\sum_{k=1}^{2j+1}\omega _k
\mid \Psi_k\rangle\langle\Psi_k\mid\,,
\end{equation}
where 
$$\omega_k\geq 0\qquad \mbox {and}\qquad \sum_{k=1}^{2j+1}\omega _k=1\,.$$
The rotated density matrix is given by the relation
\begin{equation}\label{H38}
\rho _{mm'}^{(j)}(\varphi, \theta ,\psi)=({\cal D}^{(j)}
\rho ^{(j)}{\cal D}^{\dagger\,(j)})_{mm'}\,.
\end{equation}
Here matrices ${\cal D}^{(j)}$ are the known matrices of irreducible 
representation of the group $SU(2)$ with spin $j.$ 
For the case of $j\neq 1/2,$ the formula which connects the density matrix
with marginal distribution has the natural generalized form of the spin-1/2
case:
\begin{eqnarray}\label{eq.17}
&\,&(-1)^{m_2'}\sum_{j_3=0}^{2j}\,\sum_{m_3=-j_3}^{j_3}\,(2j_3+1)^2\nonumber\\
&\times&\sum_{m_1
=-j}^{j} \int (-1)^{m_1} w\left(m_1,\varphi,\theta\right)\, D_{0 m_3}^{(j_3)}
(\varphi,\theta,\psi)\nonumber\\
&\times& \pmatrix{j&j&j_3\cr m_1&-m_1&0}\pmatrix{j&j&j_3\cr m_1'&-m_2'&m_3}
\frac{d\omega}{8\pi^2}=\rho_{m_1'm_2'}^{(j)},  \end{eqnarray}
i.e., in all the cases (where $j=1/2$ was used) 1/2 has to be replaced by 
the value of spin equal to $j.$

Thus, we obtain the generalization of the map of classical states of
an $N$-dimensional
oscillator onto a set of distributions. We formulate the result: Given $N$
coupled classical oscillators whose state is labeled by $2N$ real 
number-coordinates on its phase space
$$\mbox {\bf Q}=\left(q_1,q_2,\ldots,q_N,p_1,p_2,\ldots,p_N\right).$$
Let us construct complex $N$-vector
$$\Psi =\left(\begin{array}{c}
\Psi_1\\
\vdots\\
\Psi_N\end{array}\right),$$
where 
$$\Psi_k=\frac {1}{\sqrt 2}\left(q_k+ip_k\right);\qquad k=1,\ldots,N\,.$$
Let us consider now this vector as basis of $2j+1=N$-dimensional
irreducible representation of the group $SU(2)$ with spin $j=(N-1)/2.$
Then points {\bf Q} in the classical phase space are mapped onto the set
of probability distribution functions,
$$\mbox {\bf Q}\rightarrow w^{(j)}\left(m,\varphi,\theta\right),$$
where $m=-j,-j+1,\ldots,j-1,j$ and angles $\varphi, \theta$ 
on $S^2$.
The map is invertible up to a
common phase factor of the vector $\Psi$ which means we map one-parameter
family of classical states of $2N$-dimensional coupled oscillators
$$\left(q_k+ip_k\right)e^{i\varphi};\qquad k+1,2,\ldots,N $$
with common phase, describing a common rotation in phase space of all the
constituent oscillators, onto the set of probability distribution functions.

For $N$ coupled oscillators, one can consider the limit 
$N\rightarrow \infty ,$
which implies limit $j\rightarrow \infty .$

For spin $j,$ projections $m=-j,\ldots,j$ mean that the energy spectrum
which is mimicked by frequencies of the normal modes is an equidistant one. 
But this restriction may be easily removed.

\section{Map of Infinite-Dimensional Coupled Oscillator onto Distribution
Set and Canonical Transformations}

\noindent

We consider now the system of coupled oscillators which is obtained as a limit 
$N\rightarrow \infty$ $(j\rightarrow \infty)$ of a finite number of
coupled oscillators. To do this, we notice that to achieve the case of 
nonequidistant frequency spectrum for ``coupled classical oscillators''
corresponding to nonequidistant energy levels of a quantum system one
needs an additional map. This map will transform the wave functions of
the system with nonequidistant energy levels into wave functions of the
system with equidistant energy levels. The model of spinors is not
sensitive to the structure of energy spectrum. The information on the
energy-level position is an external one and it can be imposed if the
physical interpretation of spinor components and the corresponding 
marginal distributions are considered.

Another aspect to be mentioned is the behavior of the marginal distribution
due to canonical transformation of  position and momentum $q_k,p_k$,
i.e., of the real and imaginary parts of the wave functions.
First, let us consider linear canonical transformations
$$\mbox {\bf Q}'=\Lambda \mbox {\bf Q}\,,$$
where $\Lambda \in \mbox {Sp}\,(2N,R)$ is a symplectic real matrix.
The Poisson brackets are preserved by this transform. If the matrix
$\Lambda$ has the form
$$\Lambda =\left(\begin{array}{clcr}
\lambda_1&\lambda_2\\
\lambda_3&\lambda_4
\end{array}\right),$$
the symplectic transform can be written as
\begin{equation}\label{H39}
\Psi'=u\Psi +v\Psi^*.
\end{equation}
Here the matrix $u$ and the matrix $v$ are
$$u=\frac {\lambda_1+i\lambda_2}{2}+\frac {\lambda_4-i\lambda_3}{2}\,;
\qquad
u=\frac {\lambda_4+i\lambda_3}{2}-\frac {\lambda_1-i\lambda_2}{2}\,.
$$
Formula~(\ref{H39}) shows the action of the linear canonical transform
on wave functions in the ``classical'' formulation of quantum
mechanics~\cite{Strocchi}.

The density matrix of the pure state 
$$\rho=\Psi\Psi^\dagger$$
is transformed into 
\begin{equation}\label{H40}
\rho'=u\Psi\Psi^\dagger u^\dagger +v\Psi^*\Psi^{\mbox {tr}}v^\dagger +
v\Psi^*\Psi^\dagger u^\dagger +u\Psi\Psi^{\mbox {tr}}v^\dagger.
\end{equation}
Formula~(\ref{H40}) is an analog of Bogolyubov transform for the density 
matrix. To show this explicitly, we introduce the matrix
$$\sigma=\Psi\Psi^{\mbox {tr}}.$$
Then the transformed density matrix takes the form
$$\rho'=u\rho u+v\rho ^*v^\dagger +v\sigma^*u^\dagger +u\sigma v^\dagger.
$$
For the point canonical transformation defined by
$$\lambda_2=\lambda_3=0\,;\qquad 
\lambda _1^{-1}=\lambda _4^{\mbox {tr}},$$
 one arrives at
\begin{equation}\label{H41}
u=\frac {\lambda_1+\lambda _1^{\mbox {tr}}}{2}\,;\qquad
v=\frac {\lambda _1^{-1\,\mbox {tr}}-\lambda_1}{2}\,.
\end{equation}
Since the marginal distribution is related to the density matrix as
diagonal element of the rotated density matrix, we obtain the formula for
canonically transformed probability distribution corresponding to the
quantum state $\Psi$, namely,
\begin{eqnarray}
w'(m,\varphi,\theta)&=&\left[{\cal D}^{(j)}
u\Psi\Psi^\dagger u^\dagger {\cal D}^{(j)\dagger}
+{\cal D}^{(j)}v\Psi^*\Psi^{\mbox {tr}}
v^\dagger {\cal D}^{(j)\dagger}\right.\nonumber\\
&&\left.+{\cal D}^{(j)}v\Psi^*\Psi^\dagger u^\dagger {\cal D}^{(j)\dagger}
+{\cal D}^{(j)}u\Psi\Psi^{\mbox {tr}}v^\dagger{\cal D}^{(j)\dagger}
\right]_{mm}.\label{H42}
\end{eqnarray}
In this way, we have implemented linear canonical transformations on the 
space of distribution functions.

In infinite-dimensional case, we will use the invertible map~\cite{Mendes}
\begin{equation}\label{H43}
w_\Psi (X,\mu,\nu)=\frac {1}{2\pi|\nu|}
\left| \int \exp \,\left[\frac {i\mu y^2}{2\nu}-\frac {iyX}{\nu}\right]
\Psi(y)\,dy\right|^2
\end{equation}
of the wave function $\Psi (y)$ onto the marginal probability distribution
function $w_\Psi(X,\mu,\nu).$ The function $w_\Psi(X,\mu,\nu),$ 
with $\mu$ and $\nu$ being real parameters, is positive and normalized,
$$\int w_\Psi(X,\mu,\nu)\,dX=1
$$
for normalized wave functions,
$$ \int |\Psi(y)|^2\,dy=1\,.$$
The map can be rewritten for density matrix of pure state $\rho_\Psi(y,y')$
\begin{equation}\label{H44}
\rho _\Psi(y,y')=\Psi(y)\Psi^*(y')=x_yx_{y'}-p_yp_{y'}
+i(x_yp_{y'}-x_{y'}p_y)
\end{equation}
in the form
\begin{equation}\label{H45}
w_\Psi (X,\mu,\nu)=\frac {1}{2\pi|\nu|}
\int \rho _\Psi(y,y')
 \exp \,\left[-i\frac {y-y'}{\nu}\left(X-\mu \frac {y+y'}{2}\right)\right]
\,dy\,dy'.
\end{equation}

The Wigner--Moyal function~\cite{Wigner32,Moyal49} is related to the generic 
density matrix by
\begin{equation}\label{H46}
W(q,p)=\int \rho \left (q+\frac{u}{2},\,q
-\frac{u}{2}\right)\,e^{-ipu}\,du
\end{equation}
and
\begin{equation}\label{H47}
\rho(x,x')=\frac {1}{2\pi}\int W\left (\frac{x+x'}{2},\,p\right)\,
e^{ip(x-x')}\,dp\,.
\end{equation}
On the other hand, in the generic case the marginal distribution is related 
to the Wigner--Moyal function by
\begin{equation}\label{H48}
w\left(X,\mu,\nu\right)=\int W(q,p)
\,e^{-ik\left(X-\mu q-\nu p\right)}\frac {dq\,dp\,d\nu}{(2\pi)^2}
\end{equation}
and
\begin{equation}\label{H49}
W(q,p)=\frac {1}{2\pi}\int w\left(X,\mu,\nu\right)
e^{i\left(X-\mu q-\nu p\right)}\,dX\,d\mu\,d\nu
\end{equation}
Thus, for pure and mixed states, the probability distribution 
$w\left(X,\mu,\nu\right)$ is associated to quantum state by means of 
an invertible integral transform. This property can be used to describe
quantum states by positive probability distributions instead of wave
functions or density matrices. In the case
of classical pure state labeled by a point in the phase space 
$\left(x_q,p_q\right)$, the probability distribution 
$w\left(X,\mu,\nu\right)$ can be defined, in view of the described chain
of maps, both for finite and infinite cases.

\section{Conclusion}

\noindent

We have shown that classical states labeled by points in the classical
phase space of a classical system can be associated with the marginal 
probability
distribution functions. The map constructed uses the map of the points of
the classical phase space onto the set of wave functions of a
corresponding quantum system. Then one uses the invertible map from
wave functions onto  probability distributions introduced in the
tomography scheme for measuring the quantum states. The established relation
clarifies some aspects of the connection between quantum and classical
pictures of linear dynamical systems. This gives the possibility to transport
the results of classical considerations into quantum pictures and vice 
versa.

\section*{Acknowledgments}

\noindent

V.I.M. thanks the University of Naples ``Federico~II'' for 
kind hospitality and the Russian Foundation for Basic Research
for the partial support under Project~No.~17222.

\end{document}